\documentclass{article}
\usepackage[utf8]{inputenc}

\usepackage{listings}
\usepackage{algorithm}
\usepackage[noend]{algpseudocode}

\algrenewcommand\algorithmicindent{1.2em}
\usepackage{color, colortbl}
\definecolor{Gray}{gray}{0.9}
\newcommand{\bettersim}{{\raise.17ex\hbox{$\scriptstyle\sim$}}}
\usepackage{multirow}
\usepackage{enumitem}
\usepackage{cite}
\usepackage{graphicx}
\usepackage{url}
\usepackage{float}
\usepackage{caption}

\title{\bf{A Quick Repair Facility for Debugging}}
\author{
  Steven P. Reiss\\
  Brown University\\
  Providence, RI, USA\\
  \texttt{spr@cs.brown.edu}
  \and
  Qi Xin\\
  Wuhan University\\
  Wuhan, China\\
  \texttt{qxin@whu.edu.cn}
}
\date{}

\definecolor{darkgreen}{rgb}{0.0, 0.5, 0.0}

\begin{document}

\maketitle

\begin{abstract}

Modern development environments provide a widely used auto-correction facility for quickly repairing syntactic errors.
Auto-correction cannot deal with semantic errors,
which are much more difficult to repair.
Automated program repair techniques, designed for repairing semantic errors, are not well-suited for interactive use while debugging,
as they typically assume the existence of a high-quality test suite and take considerable time.
To bridge the gap, we developed ROSE, a tool to 
suggest quick-yet-effective repairs of semantic errors
during debugging.
ROSE does not rely on a	test suite. Instead, it	assumes
a debugger stopping point where a problem is observed.
It asks the developer to quickly describe
what is wrong, performs a light-weight fault localization to
identify potential responsible locations, and uses
a generate-and-validate strategy to produce and validate repairs.
Finally, it presents the results so the developer can 
choose and make the appropriate repair.
To assess its utility, we implemented a prototype
of ROSE that works in the Eclipse IDE and applied it to two benchmarks, QuixBugs and Defects4J, for repair. ROSE was able to suggest correct repairs for 17 QuixBugs and 16 Defects4J errors 
in seconds.
    
\end{abstract}

\maketitle

\section{Introduction}
\label{sec:intro}

Developers do software development using Integrated Development Environments (IDEs). 
Modern IDEs provide a variety of useful features including auto-correction for repairing simple errors.
A well-known tool along this line is
Eclipse's Quick Fix~\cite{quickfix}. When a compiler syntax error arises,
the developer can easily invoke Quick Fix to suggest repairs for that error, 
choose one, and make the correction.
Quick Fix is not guaranteed to work in all cases but is widely used and liked by 
developers~\cite{murphy2006java} because it can often find and repair errors quicker than the developer can.

Although useful, Quick Fix only tackles syntactic errors, those detected by a compiler.
Developers also make semantic errors
which are only exposed at run time and are much more difficult 
to repair~\cite{kim2006long,zou2015empirical}.
This motivates our research -- developing a technique that is
similar in spirit to Quick Fix but targets semantic errors
and provides quick and useful suggestions for their correction.



Automatically repairing semantic errors is the goal of automated program repair (APR) which has been well-studied for over a decade~\cite{monperrus2018automatic,goues2019automated}.  Existing APR techniques cannot be applied easily
to repairing semantic errors in an IDE because they are too slow for interactive use~\cite{liu2020efficiency}
and rely on a high-quality test suite for the code being repaired. They use the test suite for defining the problem, fault localization, and repair validation~\cite{le2011genprog}.

In practice, high-quality test suites are rare.
We scanned all Java projects in the Merobase source repository~\cite{janjic2013unabridged}. 
Of the 12,525 projects with 50 or more methods, over half (7,286 or 58\%) have no tests at all. 
Only about 13\% projects had as many tests per method as the packages in Defects4J~\cite{just2014defects4j} 
used for evaluating APR.
Even when a high-quality test suite is available,
using APR techniques that rely on a test suite can be problematic.  
Running the test suite for all potential repairs is time consuming; what is currently 
being debugged might not be a formal test case; and
recreating the environment leading to the immediate problem can be difficult.
For these reasons, addressing semantic error repair in an IDE requires novel techniques
that can work efficiently without a test suite.

We have developed a prototype system,
ROSE (Repairing Overt Semantic Errors), for repairing semantic errors while debugging Java programs in an IDE.
ROSE assumes the developer has suspended the program in the debugger and has observed a semantic problem.
ROSE first asks the developer to quickly describe what is wrong. 
Next, ROSE performs fault localization to identify locations
that might be responsible using a dynamic backward slice.
For each identified location, ROSE uses a variety of
repair generation approaches to suggest potential repairs.
%
Next ROSE validates the potential repairs without a test suite by comparing the original execution
to the repaired execution. It uses the comparison result to check if the identified problem
was fixed and to avoid 
overfitted repairs~\cite{smith2015cure}.
Finally, ROSE presents the repairs to the developer in priority order.  
The developer can choose a repair to preview 
or have ROSE make the repair.
Like Quick Fix, ROSE tries to suggest repairs quickly and does not guarantee that all problems can be 
fixed automatically.
To that end, it focuses on simple repairs.

We evaluated our prototype of ROSE,
which runs in the Eclipse IDE, on two published error benchmarks,
QuixBugs~\cite{lin2017quixbugs} and a subset of Defects4J~\cite{just2014defects4j}.
Our result shows that ROSE repaired 17 QuixBugs and 16 Defects4J errors
with a median time of suggesting a repair of about 5 seconds for QuixBugs and about 7 seconds for Defects4J errors. 
This result demonstrates the practicality of ROSE
for interactive semantic error repair.

The main contributions of this work are:
\begin{itemize}[noitemsep]
    \item A working quick-repair tool integrated into an IDE that provides 
    accurate suggestions in a reasonable time frame.
    
    \item A technique for doing fault localization based on the problem symptoms and the 
    current debugging environment.
    
    \item Techniques for validating a prospective repair without a test suite using
    comparative executions.

    \item A framework for integrating a variety of repair-generating mechanisms into an interactive error repair facility.
\end{itemize}

The rest of this paper is structured as follows. 
Section~\ref{sec:overview} provides an overview of ROSE.  
Section~\ref{sec:related_work} describes related work.  
This is 
followed by detailed descriptions of the various phases of ROSE’s 
processing (Sections~\ref{sec:problem_def} to~\ref{sec:present_result}). 
Section~\ref{sec:evaluation} describes an evaluation of ROSE with 
variations of the data sets used for evaluating APR tools.



\section{Overview of ROSE}
\label{sec:overview}

\begin{figure}[t]
\centering 
\includegraphics[width=\linewidth]{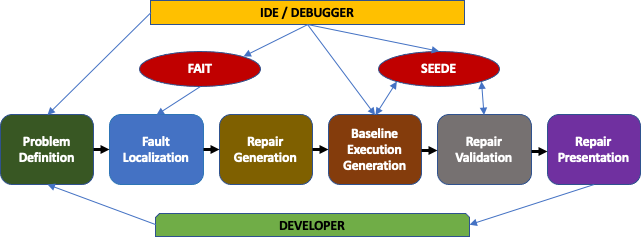}
\caption{Overview of ROSE (horizontal black arrows indicate sequencing; 
the other blue arrows represent data flow; the rounded boxes show ROSE's stages; 
the oval boxes show subsystems; and the rectangle boxes represent external environments).}
\label{fig1}
\end{figure}

Automated program repair (APR) techniques typically use a generate-and-validate approach
based on a test suite.
The techniques first perform fault localization to identify locations containing the error,  
generate potential repairs for these locations, and finally validate these repairs.
ROSE uses a similar approach, 
but does so without a test suite.
It is designed to work in conjunction with an IDE
and a debugger. An overview of the process can be seen in Figure~\ref{fig1}.

ROSE assumes the developer is using the debugger and
the program is suspended with an observed \textit{problem} caused
by a semantic \textit{error} in the program. The developer invokes ROSE at the
line where the program stopped. 
ROSE starts by asking the developer to quickly specify the problem symptoms.
It does this by querying
the debugger to get information about the
stopping point and popping up a dialog. If the program is 
stopped due to an exception, ROSE assumes that the
exception is the problem; if the program is stopped due to an assertion violation,
it assumes that is the problem.
Otherwise, the developer can indicate that a particular variable has the wrong value or that
execution should not reach this line.  The latter could arise if the code includes defensive checks 
for unexpected conditions. The problem definition process is detailed in
Section~\ref{sec:problem_def}.

After the problem is defined, the developer asks ROSE to
suggest \textit{repairs}, code changes that will fix the original problem. 
ROSE first does fault localization.
Rather than using a test suite for this purpose, ROSE uses the location where the program is suspended 
along with the problem description to identify potential lines to repair.  It does this using program 
flow analysis to compute a partial backward dynamic slice from the stopping point, which it obtains from a full dynamic
slice~\cite{agrawal1990dynamic} 
using the problem definition and execution distance to limit the result. 
This is detailed in Section~\ref{sec:error_localization}.

ROSE next generates potential repairs.
Like Quick Fix or PRF~\cite{ghanbari2020prf}, ROSE 
supports pluggable repair suggesters.
It uses pattern-based, search-based, and learning-based suggesters to quickly 
find simple, viable repairs.
For each proposed repair, these suggesters also provide a description of the repair and a syntactic priority
approximating its likelihood of being correct.
Section~\ref{sec:repair_generation} details
how ROSE generates repairs based on the suggesters.


The next step is to create a baseline execution that duplicates the original problem as a foundation for 
validating suggested repairs. 
If the execution leading to the problem is simple, for example, derived from a quick test case, 
then it can be used directly. However, in many cases, the execution is more complex 
and might have involved user or external events.
In these cases, it is more efficient and practical for ROSE to consider only the 
execution of an error-related routine on the current call stack and everything it calls.
ROSE uses SEEDE~\cite{reiss2018seede}, a live-programming facility, to create a complete execution 
history of the identified 
routine. Finally, ROSE finds the current stopping point in the execution history. 
This is described in Section~\ref{sec:baseline_execute}. 


ROSE next validates the suggested repairs using the baseline execution.  ROSE does validation 
by comparing the baseline execution of the original program with the corresponding 
execution of the repaired program, checking if the problem was fixed, and attempting 
to find a repaired run that is close to the original to minimize overfitting~\cite{d2016qlose}.
It again uses SEEDE, making the proposed repair as a 
live update and getting the resultant execution history. 
The comparison yields a semantic priority score that is used in conjunction with the 
syntactic priority derived from repair generation to create a final priority score for the repair. 
This is described in Section~\ref{sec:validate_repair}.

Finally, ROSE presents the potential repairs
to the developer as in Quick Fix and previous interactive test-suite based APR
tools~\cite{jeffrey2009bugfix,kaleeswaran2014minthint}.
It limits this presentation to repairs that are likely to be correct by showing the repairs in priority order. 
The repairs are displayed as they are validated. In this way, the developer can preview or make a 
repair as soon as it is found. Section~\ref{sec:present_result} details this.

\section{Related Work}
\label{sec:related_work}


In this section, we discuss related work in program repair, fault localization, patch overfitting, and debugging.

\textbf{Program repair}.
ROSE is closely related to automated program repair (APR) techniques~\cite{goues2019automated,monperrus2018automatic},
whose goal is to automatically fix bugs to significantly reduce manual debugging effort.
Current APR techniques usually take as input a faulty program and a test suite, which the program fails to pass. 
They typically adopt
a generate-and-validate strategy for program repair by following
three steps: (1) fault localization, (2) patch generation, and (3) patch validation.
They start by finding potential faulty locations~\cite{wong2016survey}.
Then they use various strategies to create patches 
at those locations. Finally, they validate the 
patches to identify those that yield repaired programs
passing the test suite and are promising to be correct.
Current techniques differ primarily in patch generation step
using various approaches based on 
genetic programming~\cite{le2011genprog},
repair patterns (e.g.,~\cite{liu2019tbar}),
search (e.g.,~\cite{wen2018context,jiang2018shaping,xin2017leveraging,afzal2019sosrepair}),
conditional synthesis (e.g.,~\cite{long2015staged,xiong2017precise}),
program state analysis~\cite{chen2017contract},
bytecode mutation~\cite{ghanbari2019practical},
and machine learning (e.g.,~\cite{chen2019sequencer,li2020dlfix,jiang2021cure}).
ROSE's repair generation is based on a combination of efficient prior approaches based on
patterns~\cite{liu2019tbar},
search~\cite{xin2019better}, and machine learning~\cite{chen2019sequencer}. 
We note however that it is possible to use
other patch generation approaches and consider investigating 
this as future work.

In addition to generate-and-validate-based APR techniques, ROSE is also related to 
others that are synthesis-based (e.g.,~\cite{mechtaev2016angelix,le2017s3,xuan2016nopol}),
user-interaction-based (e.g.,~\cite{bohme2020human}),
assume a non-test-suite specification
such as Alloy-based specification~\cite{gopinath2011specification},
contract~\cite{wei2010automated},
bug report~\cite{koyuncu2019ifixr},
and reference implementation~\cite{mechtaev2018semantic},
and those targeting specific faults (e.g., memory leaks~\cite{gao2015safe}
and concurrency bugs~\cite{liu2016understanding}).

\textbf{Fault localization}.
There has been a large body of research~\cite{wong2016survey} dedicated to finding program elements
that trigger a failure.
A significant portion of the fault localization techniques are 
spectrum-based~\cite{de2016spectrum}.
They assume the existence of a test suite including test cases 
that the program passes and fails to pass.
To achieve fault localization, they execute the program against
the test suite to obtain the spectra~\cite{reps1997use}
representing program's run time behavior (e.g., coverage), compare the spectra on
passed and failed test cases to compute the likelihood of an element being faulty,
and finally rank the elements based on their likelihoods.
Traditional spectrum-based techniques 
(such as~\cite{renieres2003fault,jones2005empirical,wong2013dstar})
exploit the spectra information only.
The effectiveness of these has been improved by using
mutations to investigate what elements can actually 
affect the testing results~\cite{papadakis2015metallaxis,moon2014ask},
by leveraging machine learning (e.g.,~\cite{xuan2014learning,li2019deepfl,li2021fault}),
dynamic slicing~\cite{alves2011},
and by performing causal inference (e.g.,~\cite{feyzi2019inforence,kuccuk2021improving})
to address the confounding bias issue~\cite{baah2010causal}.
In addition to spectrum-based techniques, 
there are also others that achieve fault localization via
delta-debugging~\cite{zeller1999yesterday},
value replacement~\cite{jeffrey2008fault},
predicate switching~\cite{zhang2006locating}, 
information retrieval~\cite{zhou2012should,wen2016locus,saha2013improving},
fuzzing~\cite{johnson2020causal}, and those that are based on
models (e.g.,~\cite{shchekotykhin2016efficient}).

\textbf{Patch overfitting}.
Most APR techniques rely on a test suite for validation
by checking whether a repaired program passes the whole test suite.
Because a test suite is never exhaustive in specifying all of a program's expected behaviors,
a test-suite-passing program may not correctly fix the bug.
This is known as the patch overfitting problem~\cite{smith2015cure},
which, according to the studies~\cite{qi2015analysis,motwani2020quality},
can severely harm repair quality.
Recent advances have sought to address the overfitting problem
using anti-patterns~\cite{tan2016anti},
test generation~\cite{yang2017better,xin2017identifying,xiong2018identifying},
classification~\cite{ye2021automated,wang2020automated},
patch prioritization based on rules (e.g.,~\cite{long2015staged,le2017s3}),
patch size~\cite{mechtaev2015directfix},
patch similarity~\cite{tian2020evaluating},
and machine learning~\cite{long2016automatic,saha2017elixir}.
ROSE uses a different, novel approach based on execution comparison.


\textbf{Debugging}. 
ROSE is broadly related to debugging techniques that
compute slices~\cite{agrawal1990dynamic}, identify failure-inducing inputs~\cite{zeller2002simplifying},
do interactive bug diagnosis~\cite{silva2011survey,ko2004designing}, and 
produce debugging aid work-arounds~\cite{carzaniga2013automatic}.
It differs from these techniques in producing actual repairs.
It is similar in spirit	to IDE's auto-correction facilities (such as~\cite{quickfix,autocorrect}), 
which ease debugging by automatically correcting syntax errors.
ROSE is different in being able to handle semantic errors.

\section{Problem Definition}
\label{sec:problem_def}


\begin{figure}
\centering 
\includegraphics[width=0.8\textwidth]{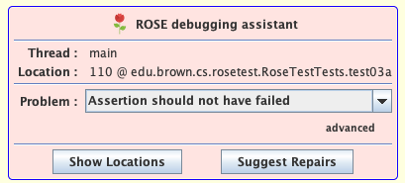}
\caption{ROSE's dialog box for problem definition.}
\label{fig2b}
\end{figure}

\begin{figure}
\centering 
\includegraphics[width=0.8\textwidth]{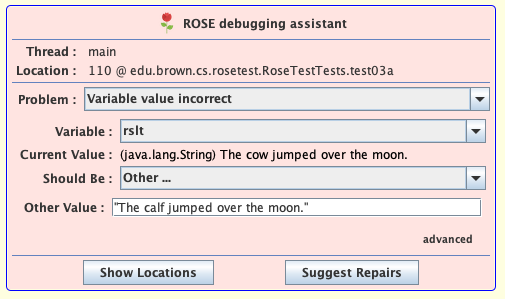}
\caption{ROSE's dialog box for variable value specification.}
\label{fig2c}
\end{figure}

ROSE assumes the program is suspended in the debugger at a
location where unexpected behavior is observed. 
The developer invokes ROSE from this stopping point by right clicking on the stopped line and selecting
``Ask ROSE to Help Debug'' from the pop-up menu.
ROSE starts by bringing
up a dialog box shown in Figure~\ref{fig2b} allowing the developer to describe the problem symptoms.

ROSE tries to make this dialog as intuitive as possible.  If the program is stopped 
due to an exception, ROSE assumes that the problem is that the exception should 
not have been thrown.  If the developer is debugging a test case 
and the program stops because a test assertion failed, ROSE assumes that the assertion 
failure is the problem.  If the program is stopped at an assert statement that failed,
ROSE assumes that is the failure.  In other cases, ROSE assumes that execution 
should not have reached the current line.  This might occur if the line represents 
defensive code that checks for an unusual or erroneous situation at which the developer 
had set a break point.  The developer has the option of approving ROSE’s choice 
or describing a different problem.  In the latter case, ROSE also lets the developer 
indicate that a particular variable has the wrong value at the point of suspension.  
It displays the current value and optionally lets the 
developer specify what the correct value should be, either specifically or 
using a constraint such as that the value should be non-null or greater than 0. 
Figure~\ref{fig2c} shows an example of this.

If the cause is due to an exception or a failed assertion, ROSE further determines the expression 
that caused the problem to assist in fault localization.  It does this 
using a combination of code analysis of the line in question and querying values from 
debugger for subexpressions.  It currently handles exceptions caused by assert statements, 
JUnit assertions, null pointers, index out of bounds for arrays, lists and strings, and 
failing casts. For example, for a NullPointerException, ROSE will look at all 
subexpressions of the executing statement in evaluation order, and for each that 
might generate a null pointer exception, uses the debugger 
to evaluate the subexpression to see if it is currently null.

Once the developer has defined the problem symptoms, she can use the dialog 
to ask ROSE to suggest appropriate repairs by clicking ``Suggest Repairs''. Again, for suggesting repairs, 
ROSE only needs a description of the problem symptoms. The developer does not need to know the error or its location.

\section{Fault Localization}
\label{sec:error_localization}

\begin{algorithm}[H]
  \caption{Fault localization algorithm sketch.}
  \label{fig3}
  \tiny
\begin{algorithmic}[1]
    
\Procedure{FindFaultyLocations}{Location loc,Reference r}
  \If {r == null} ctx = [ r->10, COND->10 ]
  \Else \ ctx = [ COND->10 ]
  \EndIf
  \State queue.add(<loc,ctx>) \Comment{queue is a global variable}
  \State \Call{ProcessQueue}{queue}
\EndProcedure

\Procedure{ProcessQueue}{queue}
  \While {queue is not empty}
      \State <loc,ctx> = queue.pop()
      \State \Call{ComputeNext}{loc,ctx}
  \EndWhile
\EndProcedure

\Procedure{ComputeNext}{Location loc,Context ctx}
  \If {loc has been processed before}
      \State ctx = merge ctx with prior context
      \If {ctx == prior context} return
      \EndIf
  \EndIf
  \If {loc is start of a method}
      \State callctx = new call context any locals in ctx mapped to arguments on stack at call site
      \State Decrement callctx COND priority
      \For {each call site of this method consistent with stack}
          \State Let callloc = location of call \State queue.add(<calloc,callctx>)
      \EndFor
  \Else
      \For {each prior location priocloc of loc}
          \If {priorloc is consistent with current values}
            \State \Call{HandleFlowFrom}{priorloc,ctx,loc}
           \EndIf
      \EndFor
  \EndIf
\EndProcedure

\Procedure{HandleFlowFrom}{Location priorloc, Context curctx, Location curloc}
  \State BackflowData bfd = \Call{computePriorStateContext}{curctx,curloc,priorloc}
  \State Context priorctx = bfd.prior\_context
  \If {loc is a method call} add relevant arguments to priorctx
  \EndIf
  \For {each AuxReference aref in bfd.aux\_references}
      \State refctx = create context with [aref->Max(priority in curctx)-1]
      \If {refctx is relevant} queue.add(<aref.location,refctx>)
      \EndIf
  \EndFor
  \If {priorloc is a method call and method is relevant}
      \For {each method m called from priorloc}
         \State returnctx = new context containing return value and COND
         \State queue.add(<return location of m,returnctx>)
      \EndFor
  \EndIf
  \If {priorctx is relevant} \State queue.add(<priorloc,priorctx>)
  \EndIf
\EndProcedure

\Procedure{ComputePriorStateContext}{Context ctx, Location curloc, Location priorloc}
   \State Let newctx = [ ], arefs = \{ \}
   \For {each <ref->priority> in ctx}
      \State BackFlow bf = \Call{ComputeBackFlow}{priorloc,curloc,ref}
      \If {bf.reference != null} add [bf.reference->priority] to newctx
      \EndIf
      \For{each AuxReference aref in bf.aux\_references}
         \If {aref.location == priorloc} Add    [aref.reference->priority-1] to newctx
         \Else \ add aref to arefs
         \EndIf
      \EndFor
   \EndFor
   \If {ctx contains COND->priority} 
      \State BackFlow bf = \Call{ComputeBackFlow}{priorloc,curloc,null}
      \For {each AuxReference aref in bf.aux\_references}
         \If {aref.location == priorloc} Add [aref.reference->priority-1] to newctx
         \Else \ add aref to arefs
         \EndIf
      \EndFor
   \EndIf
   \State return new BackFlowData with newctx and arefs
\EndProcedure

\Procedure{ComputeBackFlow}{Location fromloc,Location toloc,Reference ref}
   \State Let newref = null, arefs = \{ \}
   \State Consider the code executed between fromloc and toloc
   \If {ref is computed by the code}
      \If {ref is loaded from local/field/array} newref = reference to source of load
      \Else \ add any parameters of the computation as arefs
      \EndIf
   \ElsIf {ref is a stack reference} newref = updated stack reference based on stack delta of instruction
   \Else \ newref = ref
   \EndIf
   \If {ref is empty and code is a conditional branch}
      \State add any parameters used on condition as arefs
   \EndIf
   \State return new Backflow with newref and arefs
\EndProcedure


\end{algorithmic}
\end{algorithm}

ROSE attempts to find potential lines containing the error without a test suite. 
It starts with the 
assumption that the developer is debugging their system and has stopped at a point that 
is relevant to the error.  It identifies candidate lines that could directly cause the identified problem symptoms.
It assigns priority to these lines based on the execution distance (essentially the distance in a program dependence graph)
between the line and the identified 
symptom.  It does this by computing a partial backward slice specific to the problem 
at the developer’s stopping point using values from the debugger and the call stack. 

ROSE uses the interactive FAIT flow analysis tool~\cite{reiss2019continuous}.  FAIT updates control and data flow 
information about the system as the programmer types using abstract interpretation~\cite{cousot1997program}.  
This associates an abstract program state (stack, variable, and memory values) with each execution
point.  FAIT provides a query
mechanism that can start at a particular location with a particular variable or program state 
and show how that variable or state could have been computed.

We modified the FAIT query mechanism to provide a list of lines that might 
affect a particular value at a particular execution location and their execution distance.  
This lets ROSE find
lines that can cause an exception or a bad variable value and lines that cause 
flow to reach the stopping point.  An abstract of this algorithm is shown in Algorithm~\ref{fig3}.
Using flow analysis to identify potential error locations lets ROSE take advantage of 
knowledge of the problem and access to current values in the debugger environment. 
The basic idea is to associate a set of relevant references with each program 
state and to assign a priority to each reference while maintaining the set of 
known values.  A reference denotes a stack element, a local variable, a field, or 
an array element that contains a value relevant to the problem; the priority notes 
the execution distance (assignments or computations) of the reference from the stopping point.  
A special reference keeps track of whether control flow is relevant.  
The set of references and known values is contained in a \textit{Context}; 
the point in the program, either a particular byte code instruction or an 
abstract syntax tree location, is contained in a \textit{Location}.  

The initial set of program references is determined by the problem symptom.  For example, a problem 
based on a variable includes a reference to the variable, and a problem based on an exception 
includes a reference to the stack element containing the expression value causing the exception.  
The initial context also indicates that control flow is relevant. The set of initial values 
is obtained from the debugger. The algorithm uses a 
work queue mechanism to process each location and associated context (lines 7-9).

For a given location and context, the algorithm needs to compute any prior locations, 
those that would be executed immediately before the current one, and determine the 
proper context for that location based on the current context.  This computation is 
based on what code was executed between the prior and the current location.  If 
the current location corresponds to the start of a method, then the algorithm considers 
all callers to that method in the static analysis that match the current call stack 
and sets up a new context at each call site, mapping any local parameters to the arguments at the call, 
and decrementing the condition count (lines 14-19).
Otherwise, it uses the flow analysis to determine prior program points where the abstract
values are consistent with any current known values and considers each in turn (lines 20-23).

The algorithm considers what was executed between the prior state context and the current one.  
FAIT includes code to determine, for a given reference, both what that reference was at 
the prior state and what other references might have been used to compute it (lines 52-61).  
A stack reference might change its location in the stack, might disappear, or 
might become a reference to a local variable (on a load); a local reference might be 
changed to a stack reference (on a store).  A field load adds all locations where that field 
was stored as auxiliary references; an array load adds all locations where array elements were set. 
A stack reference computed using an operator, 
adds the operands as auxiliary references.  
For example, if the top of the stack was relevant and the code was an ADD, then
the reference to the top of the stack would be removed, but auxiliary references to the top two
stack elements (the operands of the ADD) would be added with a decremented priority.
This information is combined for each reference in the context (lines 39-44) and for 
control flow if relevant (lines 45-49).  The resultant contexts and locations are then queued for future
consideration (lines 28-30, 35-36).

Method calls are handled explicitly by the algorithm.  
A method is relevant if its return value or the `this' parameter is relevant.  If the call is
relevant, the algorithm  adds all the parameters to the prior state (line 27).  If control flow is relevant, 
it also queues the return site of any method called at this point with a context that 
includes the return value if that was relevant (lines 31-34).

The result is a set of program points where a relevant value or relevant control flow was computed.  
This is the set of potential error locations to be considered.  
The priority of the location is determined by the maximum priority of an item in the context.   
From this set, ROSE computes the set of potential error lines.  
It can optionally exclude from this set any test routines or drivers to ensure  
that ROSE fixes the problem rather than changing the test.

We note that this type of analysis yields different results than control-flow 
or spectra-based methods. Spectra-based localization, because it uses coverage information, 
only provides information at the basic-block level.  ROSE identifies specific expressions 
that can directly affect the problem.  Statements that are not indicated as erroneous by ROSE 
do not affect the problem.  On the other hand, locations where a new statement should be 
inserted in a particular branch can be identified by spectra-based techniques but will be 
ignored by ROSE.  ROSE assumes that the latter case can be deferred to repair 
generation.

\section{Repair Generation}
\label{sec:repair_generation}


The next step is to generate potential repairs for each identified potential error location.  Like 
Quick Fix in Eclipse or PRF~\cite{ghanbari2020prf}, ROSE supports pluggable repair suggesters. 
Because it is designed 
to operate within an IDE, ROSE concentrates on suggesters that are fast and  
do not generate too many spurious potential solutions.  The suggesters are 
designed to yield a repair, a description of that repair for the developer, 
and an estimate of the probability of the repair being correct (its syntactic priority).  
ROSE's current suggesters are based on existing APR 
techniques.  The suggesters are prioritized so that those more likely to 
find correct solutions quickly are considered first.  

The suggester with the highest priority looks for code smells~\cite{tufano2015and}, 
i.e., common programming errors, 
that may occur at a potential error line.
This is a pattern-based analysis where the set of patterns is based on various studies of 
common errors~\cite{bryce2010one,catolino2019not} and bug fixes~\cite{pan2009toward,campos2017common} as well as our own 
programming experience. Current patterns include ones dealing with equality such as using \textit{==}
for \textit{=},
using \textit{==} rather than \textit{equals}, and using \textit{==} for an assignment;
ones dealing with strings such as using
\textit{toString} rather that \textit{String.valueOf} and calling methods like \textit{String.trim} but not
using the result; ones dealing with operators often used incorrectly such as exclusive or or complement;
and ones dealing with confusing integer and real arithmetic.
The suggester with
the second highest priority is one that handles avoiding exceptions by inserting or modifying 
conditionals around the code where an exception was thrown, a common fix.
Most of the remaining pattern-based suggesters currently have the same priority.  These include one 
that handles conditionals by changing the relational operator; one that considers different 
parameter orders in calls; one that looks for common errors with loop indices in \textit{for} 
statements; and one that considers replacing variables or methods with other accessible variables 
or methods with the same data type.  

In addition, we have included suggesters that use existing APR tools. 
One uses sharpFix~\cite{xin2019better}, a code-search based tool, to find suggested repairs 
based on similar code in the current project.
This has slightly lower priority since it is slower than the simple, pattern-based suggesters, 
but has proven effective.  The results from sharpFix are reordered and filtered to avoid
duplicating the other suggesters. 
Another external suggester uses SequenceR~\cite{chen2019sequencer}, a machine-learning based tool, to produce repairs. 
ROSE again
filters and sorts the results to avoid duplication.
The current implementation of SequenceR is relatively slow
and is prone to generating
repairs that do not compile, so this suggester
has the lowest priority.
As we will show in Section~\ref{sec:evaluation}, this combination of 
repair generation techniques is effective. 
Rose could use other repair generation approaches~\cite{monperrus:hal-01956501} 
to improve the recommendations. We expect to investigate these and modify priorities
as the system evolves.

To generate repairs, the system runs each suggester on each potential error line.
These runs are prioritized using both the suggester and location priorities, and are run 
in parallel with multiple threads using a thread pool with a priority queue. 

\section{Generating a Baseline Execution}
\label{sec:baseline_execute}

Once potential repairs have been generated, they need to be validated.  
In the absence of a test suite, ROSE does this by comparing the repaired execution with the 
original execution to see if the problem symptoms go away, the repair is likely to be 
non-overfitting~\cite{smith2015cure}, and the program seems to work.
The first step in this process is to create a baseline execution that shows how the program 
got into the current state based on the current debugging environment.  
This is done using the live programming tool SEEDE.

SEEDE~\cite{reiss2018seede} is a practical live-programming facility that works in 
collaboration with the debugger to simulate the execution of a target method and everything 
it calls and update the execution after each edit. SEEDE generates an execution trace that includes
the history of
each variable, field, array element that is updated including the 
original and new values and the time stamp. The trace also includes a time stamp for
each executed line and function call. 


ROSE creates a SEEDE execution that includes the stopping point and potential repairs.  
ROSE is designed to handle a general debugging situation and does not assume a test suite or 
even a test case.  ROSE 
could be invoked from the middle of a run that involved user or other external 
interaction.  
It could be invoked from a run that has gone on for several minutes or hours.  In many of these cases, 
it is not practical
to recreate the current debugging environment from an initial execution.  Instead, ROSE tries
to do a local partial reexecution starting with a routine on the current execution 
stack using the
current environment in order to duplicate the problem.  This approach is also more 
efficient for validating potential repairs as each validation will involve less computation.  
To find such a partial reexecution, ROSE needs to determine the routine 
at which to begin and needs to ensure that the simulated execution from SEEDE matches 
the actual execution.

To determine where to begin execution, ROSE first identifies the closest stack frame that 
includes all potential locations identified by fault localization directly or indirectly.
This is relevant when the stopping point is in one routine, but the error is in one of the 
callers of that routine or in a routine one of the callers invoked.  Next, ROSE considers whether 
it is worthwhile to consider a caller of this routine rather than the routine itself.  
This can be beneficial for several reasons.  First, it can provide additional information 
for testing purposes, for example if the erroneous routine is called multiple times with different 
arguments or if the erroneous routine might return a value that would later cause an error.  
Second, it can be helpful in ensuring the simulated environment matches the actual execution 
by forcing the reinitialization of values that were changed in the actual execution.  
Currently, ROSE will consider a test driver, main program, or user callback from a system routine 
if it is close to the selected routine.  The user interface ROSE provides also lets the developer 
specify the routine if desired.

Ensuring the simulated execution matches the actual execution can be challenging when the code 
changed a parameter or global variable the execution depends on.
In general, it is not 
possible to undo the execution back to the chosen starting point.  However, 
this is often feasible in practice.  A typical case is where the program checks whether a value has 
previously been computed by trying to add it to a \textit{done} set, and just returns if it has.  
If the actual execution added a value to that set, then the simulated execution would just return and 
would not match the actual one.  SEEDE provides a mechanism for defining additional initializations 
before creating the execution which ROSE uses in an attempt to tackle this problem.

If the executions do not match, ROSE does a simple backward program analysis from the 
stopping point in the selected method to determine the set of variables that might have changed.  
This is then restricted to variables or fields that are external to the given method.  
If these were parameters with simple values, ROSE will compute 
the values from the caller by evaluating the corresponding passed expression in the debugger.  If the 
value is more complex, for example the \textit{done} set mentioned above, ROSE tries a set of 
heuristics to create appropriate initializations for SEEDE.  This is done using a SEEDE 
run that fails to match the original execution. ROSE finds the first point in this 
run where the execution might differ and considers what evaluations had been done up 
to that point on the identified set of variables.  
If it finds something added to a map or collection, it will try removing it; if it finds something removed from a 
collection or map, it will try adding it.  Each potential initialization is tried 
to see if matched execution proceeds further.   
If the simulated and actual runs cannot be matched, then ROSE will not be able to 
validate any repair and it yields no results.

Choosing a higher-level starting point involves a trade-off between effectiveness and efficiency.  
Higher-level starting points will often initialize the problematic values directly, thus eliminating the need for 
the above analysis.  
This is why ROSE prefers starting at a test case, main program, or callback from a system routine.
However, a higher-level starting point can greatly increase the validation time, making ROSE impractical for 
interactive use.  


\section{Validating a Repair}
\label{sec:validate_repair}

\begin{algorithm}[H]
  \caption{Repair validation algorithm.}
  \label{fig4}
  \scriptsize
\begin{algorithmic}[1]

\Procedure{Validate}{Execution orig,Execution repair}
   \If {repair has a compiler or run time error} return 0
   \EndIf
   \State Matcher matcher = \Call{ComputeMatch}{orig,repair}
   \If {!matcher.executionChanged()} return 0
   \EndIf
   \State double score = \Call{Validate<ProblemType>}{matcher}
   \State double closeness = difference\_time / problem\_time
   \State return score * 0.95 + closeness * 0.05
   \State 
\EndProcedure

\Procedure{ValidateException}{Matcher matcher}
   \If {call of program point can be found in repair} 
      \If {repair doesn't throw exeception} 
          \If {program point can be found in repair} return 1.0
          \Else $ $ return 0.8
          \EndIf
      \ElsIf {exception thrown after match point} return 0.5
      \ElsIf {different exception thrown} return 0.1   
      \Else $ $ return 0
      \EndIf
   \EndIf
   \If {top method returned} return 0.75
   \EndIf
   \State return 0.2 
\EndProcedure

\Procedure{ValidateAssertion}{Matcher matcher}
    \State return ValidateException(matcher)
\EndProcedure    

\Procedure{ValidateVariable}{Matcher matcher}
   \If {exception thrown} return 0 
   \EndIf
   \If {match found}
      \If {value at match == original value} return 0
      \EndIf
      \If {target value given}
         \If {value at match matches target value} return 1
         \Else \ return 0
         \EndIf
      \Else \ return 1 
      \EndIf
   \Else
      \State boolean haveold = variable ever held original value
      \State boolean haveother = variable held other value (after old value)
      \If {target value given}
         \If {value ever matches target value}
            \If {haveold} return 0.6 
            \Else \ return 0.75
            \EndIf
        \EndIf
      \ElsIf {!haveold} return 0.6 
      \ElsIf {haveother} return 0.5 
      \EndIf
      \State return 0.0
   \EndIf
\EndProcedure      

\Procedure{ValidateLocation}{Matcher matcher}
   \If {no match found and matching call found}
      \If {exception thrown} return 0.2 
      \Else \ return 0.8 
      \EndIf      
   \EndIf
   \If {matching call not found}
      return 0.2
   \EndIf
   \State Return 0
\EndProcedure

\end{algorithmic}
\end{algorithm}

ROSE validates a repair by comparing the baseline execution with the repaired execution from the same 
starting point using the same initializations.
To obtain the repaired execution, ROSE uses SEEDE to reexecute the code after the repair has been made.
This yields a full execution
trace, which ROSE uses to compare with the baseline execution trace. The comparison yields a validation 
score between 0 and 1, with 0 indicating that the repair is invalid and 1 
indicating the highest degree of confidence in the repair.  A sketch of the algorithm used is shown in Algorithm~\ref{fig4}.

ROSE considers the suggested repairs, looking at them in order based on their associated syntactic priority.
Multiple repairs are considered in parallel using multiple threads. ROSE keeps track of the number of 
repairs it has validated and the total execution cost of validating those repairs.  It also tracks 
whether any repair was considered likely to work (score >= 0.7).  It stops considering repairs when 
either a maximum number of  repairs have been considered or when the total 
cost (number of steps of SEEDE execution)
exceeds a limit.
These bounds are dependent on whether there has been a likely repair and are designed 
so that ROSE finishes within a reasonable interactive time frame
even if it means not reporting a repair that might otherwise be found.

If the proposed repair either does not compile or yields a run-time type 
error, then the repair is considered invalid (line 2). 
Otherwise, the repaired 
execution is matched with the baseline execution by \textit{ComputeMatch} (line 3). This comparison goes through the two
execution traces from SEEDE step-by-step, checking both control flow and data values. It finds the first 
location where the control flows differ and the first location where data values differ.  It 
also finds the call in the repaired execution corresponding to the call in which 
the program stopped, and the point in the repaired execution that corresponds 
to the stopping point in the original execution.  

Because repair locations are computed using static analysis, it can be the case that 
the repaired execution matches the baseline execution, at least up to the problem point.
When that happens, ROSE
considers the repair invalid as it had no effect (line 4).
Otherwise, ROSE validates the repaired execution based on the initial definition
of the problem, which can be related to exceptions, assertions, variable values, or locations.

If the problem involves either an unexpected exception or an assertion violation, 
then the validator checks if the repaired run also threw an exception and scores the repair to find an 
execution close to the original that does not throw an exception, 
particularly, the original exception.
If the repaired execution includes the problem context, then if no exception was thrown, 
the repair is considered valid if the problem line was executed (line 12) or mostly valid if it was not (line 13).
If an exception was thrown after the matching point, then the repair is considered possibly valid (line 14);
if a different exception was thrown the repair is considered unlikely (line 15); if the same exception
was thrown the repair is considered invalid (line 16).  If there was no matching problem context, the
repair is considered probably valid if no exception was thrown (line 17) and probably invalid if one was (line 18).

If the problem involves a variable value, 
ROSE uses a different approach.
In this case, ROSE scores the repairs by finding a repaired execution that is close to the
baseline where the variable 
takes on either the target value consistent with the value
specified by the developer when defining the problem or a value different from the original if 
the target is not known. If an exception is thrown in the repaired execution, it is 
considered invalid (line 22).  Otherwise, if the program point can be matched in the repaired execution, 
the value of the variable is considered at that point.  If the repaired value is the 
same as the original, the repair is considered invalid (line 24).  Otherwise, the repair is 
considered valid if the repaired value is consistent with the target value and invalid if not (line 26-28). 
If no match for the program point can be found, the values the variable in question takes on are
considered.  If the variable takes on the original value and not something consistent
with the target, it is considered invalid (line 38).  Otherwise, the repair is considered probably 
valid with the score depending on whether the original value was taken on and whether a value 
consistent with the problem specification was taken on and in what order (lines 32-37).
If the problem involves reaching an unexpected line, 
the score is computed based on whether the line was executed. ROSE penalizes executions that 
avoid the line by throwing 
an exception or by not calling the problem method in the first place.  If the program 
point can be matched in the repaired execution, and the line is still executed, the 
repair is considered invalid (line 44).
If the repaired code entered the same method but did not execute the line, the repair
is considered likely to be correct if no exception is thrown (line 42) and unlikely if
one is (line 41); otherwise the repair is considered unlikely (line 43).

ROSE adjusts all these scores to favor results where the change occurs later in 
the execution.  This is done by considering the point where the execution 
(either data or control flow) changes relative to the problem point and incorporating 
this fraction as a small part of the score (lines 5-7).  
ROSE later adjusts the score returned by this algorithm by checking if the partial execution 
completed by returning or if it failed by throwing an exception or infinite looping.

In Algorithm~\ref{fig4}, ROSE uses different constants to approximate the likelihood of a repair's correctness 
in various situations. We determined these constants based on an initial suite of problems for testing ROSE 
and on our extensive debugging experience. ROSE's validation result is not sensitive to 
minor changes in these constants.

\section{Presenting the Result}
\label{sec:present_result}

\begin{figure}
\centering 
\includegraphics[width=0.8\textwidth]{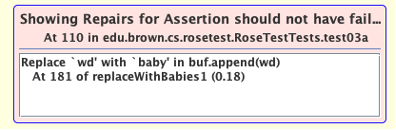}
\caption{Suggested repairs.}
\label{fig2e}
\end{figure}


Once a repair has been validated, it can be sent to the front end to be displayed to the developer. 
The developer can choose to preview or make the repair as soon as it is displayed. 
ROSE maintains a validity threshold
and only sends repairs whose scores are above this.  The front end computes a final repair probability by
combining the syntactic priority
computed by repair generation and the semantic priority computed by validation.  It then provides
a view of the current repairs ordered by this combined score, as shown in Figure~\ref{fig2e}.  
This view is updated as repairs are found.  This provides a live, on-line 
view of the potential repairs found by ROSE.  

The developer can interact with the repair window.  They can choose a particular 
repair and ask for a preview.  This brings up a window showing the code before and after the repair
with the differences highlighted.  Alternatively, the 
developer can ask ROSE to make the repair.  In this case, using the hot-swap capability of 
Java, the developer can continue execution with the repaired code.

\section{Evaluation}
\label{sec:evaluation}




We implemented a prototype of ROSE that worked with Eclipse and chose two widely used benchmarks for
APR~\cite{ye2021comprehensive, martinez2017automatic}, QuixBugs and Defects4J (v2.0.0), to evaluate its repair efficacy.
QuixBugs~\cite{lin2017quixbugs} is a set of 40 student programs containing typical programming errors. 
Defects4J~\cite{just2014defects4j} is a corpus of more complex development errors. 

Our evaluation attempts to answer the question: \emph{Is it possible to
do program repair interactively while debugging without a test suite?}
To do this we looked at 
how many errors can be 
repaired and how quickly can this be done.
As ROSE resembles existing APR tools, 
we evaluated it along similar lines. 
We note that this is not a direct comparison as the inputs, outputs, 
and goals of ROSE differ from those of traditional program repair tools.  


\subsection{Experimental Setup}

For the QuixBugs benchmark, we created an Eclipse workspace containing all 40 
unrepaired programs but no tests.  For each unrepaired program,
we created a main program that effectively ran the failing test.
The result of the test was checked using either the original JUnit assertion 
or a simple assert statement.  For each main program, we created a launch configuration and 
ran it.  If the program stopped with an exception or an assertion violation, we used that as 
the symptoms. If the program infinitely looped, we set a break point in the loop, had the 
execution stop at the break point after several 
times through the loop, and then identified 
either an incorrect statement or variable value as the symptom.  We then
ran ROSE on each of the 40 programs to generate repairs. 
Following a standard way to evaluate automated repairs, we considered a repair correct 
if it was semantically the same as that provided in the benchmark (the ground truth repair). 
While determining semantic equivalence can be challenging, it was clear and easy in these cases.

For the Defects4J benchmark, we selected a set of errors (bugs) to repair that were
relevant to the assumptions of ROSE, where the correct repair was simple and
close to the stopping point.  We determined simplicity by ensuring the repair 
only involved a single line being modified or added.
We determined if the repair was close to the stopping point by considering how much execution
was needed from a logical starting point to the stopping point.  We wanted this to be
fewer than 500,000 instructions and either chose a more appropriate starting point or discarded
the error if this could not be achieved.    
Our initial set of errors was those with one-line fixes used to evaluate 
SequenceR~\cite{chen2019sequencer}. We identified different projects, Chart, Closure, Lang, Math, and 
Time, within this and set up an Eclipse workspace for each. For Math, we set up two workspaces, 
Math and Math3, since the base directory 
changed over time.   
For each workspace, we used the source from the most recent error after discarding all test cases.  
Then we recreated the remaining errors in that source.  We discarded
errors where this was not possible.  Next, we created a main program for each error
based on a failing test, replacing the JUnit assertion of the test with an assert 
statement to avoid requiring the JUnit library.  We created a separate 
launch configuration for each error
and ran it.  We used the failing assertion or a thrown exception as the problem symptom 
for ROSE.  A few of the tests were discarded as they no longer failed.  

Some of the cases 
involved running a large body of code to elicit the problem.  ROSE assumes that the developer 
has done preliminary debugging and has stopped relatively close to the problem.  
For two errors, Math30 and Math82, we set an internal break point.  
We discarded all the Closure errors as they involved running a 
compiler on a segment of code, printing the result, and comparing the output to the 
expected output.  This involved many source files and considerable execution and was 
outside the targeted purpose of ROSE.  
We ended up with 32 different Defects4J
errors shown in Table~\ref{table2}.

Next, as with QuixBugs, we ran ROSE on each of the errors in each workspace.  We considered a repair 
suggested by ROSE to be correct if it was the same or semantically equivalent to the Defects4J 
corrected code at the same line.

It is worth noting again that ROSE does not assume the existence of a test suite or even a test case. 
In this evaluation, we used failing test cases only to identify the problem to fix so the 
results could be compared to prior work.

The QuixBugs benchmark~\cite{lin2017quixbugs} is a reflection of typical development errors made by students.
The Defects4J benchmark includes a variety of different types of errors and mainly errors 
that persisted over major check-ins.  The ones we selected are 
typical of the simpler problems that occur during development and debugging~\cite{bryce2010one,catolino2019not}.
Thus the errors used in this evaluation are representative of those encountered during software development. 
As a further demonstration of this, we surveyed the last 3 months of commits 
of the large system 
Code Bubbles~\cite{bragdon2010code} and showed that about 30\% of 
bug fixes during development
are the simple kind that are targeted by ROSE.

\subsection{Results}

\begin{table}[H]
\centering
\scriptsize
\caption{Results for QuixBugs errors.}
\resizebox{\textwidth}{!}{
\begin{tabular}{|c|l|l|l|l|l|l|l|l|} \hline
\textbf{Correctly Repaired?} &
\textbf{Error Names / Statistics} & \textbf{\#Results} & \textbf{Correct Rank} & \textbf{Total Time} & \textbf{Fix Time} &
\textbf{Fix Count} & \textbf{SEEDE Count} & \textbf{\#Checked} \\ \hline \hline

\multirow{20}{*}{Yes}
& BitCount & 1 & 1 & 1.1 & <0.1 & 1 & 68 & 51 \\
& BreadthFirstSearch & 6 & 1 & 6.6 & 1.4 & 11 & 42508 & 124 \\
& BucketSort & 1 & 1 & 3 & 0.7 & 8 & 21733 & 32 \\
& DepthFirstSearch & 4 & 4 & 3.6 & 0.8 & 6 & 10169 & 75 \\
& DetectCycle & 9 & 1 & 2.7 & <0.1 & 4 & 390 & 71 \\
& FindFirstInSorted & 1 & 1 & 6.2 & 0.7 & 6 & 60257 & 101 \\
& Gcd & 1 & 1 & 1.8 & 0.1 & 1 & 97 & 35 \\
& Hanoi & 3 & 1 & 4.9 & 0.5 & 14 & 9655 & 165 \\
& Knapsack & 1 & 1 & 5.6 & 1 & 20 & 67129 & 167 \\
& LongestCommonSubsequence & 2 & 2 & 7.1 & 3.7 & 29 & 373139 & 83 \\
& NextPermutation & 2 & 2 & 28.2 & 0.1 & 3 & 3559 & 143 \\
& Pascal & 1 & 1 & 10.9 & 0.8 & 22 & 52394 & 131 \\
& PossibleChange & 1 & 1 & 2 & 0.1 & 5 & 3120 & 24 \\
& QuickSort & 2 & 1 & 5.6 & 0.9 & 15 & 157357 & 182 \\
& RpnEval & 1 & 1 & 5.3 & 0.9 & 11 & 8771 & 107 \\
& Sieve & 1 & 1 & 3.7 & 1.1 & 11 & 45542 & 51 \\
& TopologicalOrdering & 2 & 1 & 23.6 & 2.4 & 24 & 74505 & 157 \\ \cline{2-9}
& \textbf{Median} & 1 & 1 & 5.3 & 0.8 & 11 & 21733 & 101 \\
& \textbf{Average} & 2.3 & 1.3 & 7.2 & 1 & 11.2 & 54729 & 99.9 \\
& \textbf{StdDev} & 2.2 & 0.8 & 7.5 & 0.9 & 8.4 & 91568.7 & 52 \\ \hline \hline

\multirow{23}{*}{No}
& FindInSorted & 27 & -1 & 2.9 & 0 & 0 & 0 & 98 \\
& Flatten & 0 & -1 & 2.1 & 0 & 0 & 0 & 19 \\
& GetFactors & 2 & -1 & 3.5 & 0 & 0 & 0 & 80 \\
& IsValidParenthesization & 4 & -1 & 2.3 & 0 & 0 & 0 & 52 \\
& KHeapSort & 0 & -1 & 3 & 0 & 0 & 0 & 54 \\
& Kth & 0 & -1 & 8.1 & 0 & 0 & 0 & 3 \\
& LcsLength & 1 & -1 & 9 & 0 & 0 & 0 & 135 \\
& Levenshtein & 0 & -1 & 12.4 & 0 & 0 & 0 & 30 \\
& Lis & 8 & -1 & 11 & 0 & 0 & 0 & 129 \\
& MaxSublistSum & 0 & -1 & 1.8 & 0 & 0 & 0 & 31 \\
& MergeSort & 0 & -1 & 1.9 & 0 & 0 & 0 & 15 \\
& MinimumSpanningTree & 0 & -1 & 4.1 & 0 & 0 & 0 & 96 \\
& NextPalindrome & 0 & -1 & 4 & 0 & 0 & 0 & 90 \\
& PowerSet & 0 & -1 & 20.8 & 0 & 0 & 0 & 82 \\
& ReverseLinkedList & 1 & -1 & 5.1 & 0 & 0 & 0 & 89 \\
& ShortestPathLength & 0 & -1 & 16.2 & 0 & 0 & 0 & 413 \\
& ShortestPathLengths & 1 & -1 & 7.4 & 0 & 0 & 0 & 53 \\
& ShortestPaths & 0 & -1 & 8.7 & 0 & 0 & 0 & 154 \\
& ShuntingYard & 0 & -1 & 4.6 & 0 & 0 & 0 & 89 \\
& Sqrt & 0 & -1 & 3.4 & 0 & 0 & 0 & 30 \\
& Subsequences & 0 & -1 & 5.2 & 0 & 0 & 0 & 122 \\
& ToBase & 0 & -1 & 3.8 & 0 & 0 & 0 & 40 \\
& Wrap & 2 & -1 & 3.9 & 0 & 0 & 0 & 66 \\ \hline

\end{tabular}}
\label{table1}
\end{table}

\begin{table}[H]
\centering
\scriptsize
\caption{Results for Defects4J errors.}
\resizebox{\textwidth}{!}{
\begin{tabular}{|c|l|l|l|l|l|l|l|l|} \hline
\textbf{Correctly Repaired?} & \textbf{Error Names / Statistics} & \textbf{\#Results} & \textbf{Correct Rank} & \textbf{Total Time} & \textbf{Fix Time} & \textbf{Fix Count} & \textbf{SEEDE Count} & \textbf{\#Checked} \\ \hline \hline

\multirow{19}{*}{Yes}
& Chart01 & 1 & 1 & 39.7 & 10.3 & 30 & 105152 & 13 \\
& Chart11 & 2 & 2 & 31.1 & 8 & 35 & 35720 & 210 \\
& Chart12 & 1 & 1 & 44.9 & 6.8 & 17 & 315938 & 124 \\
& Chart20 & 1 & 1 & 8.7 & 0.5 & 2 & 393 & 14 \\
& Lang06 & 2 & 1 & 12.1 & 7.3 & 31 & 355343 & 66 \\
& Lang33 & 3 & 2 & 7.2 & 2.3 & 4 & 149 & 45 \\
& Lang59 & 1 & 1 & 7.8 & 3.2 & 33 & 1316 & 77 \\
& Math59 & 1 & 1 & 18.1 & 7.9 & 9 & 263 & 15 \\
& Math69 & 2 & 1 & 48.6 & 41.6 & 143 & 1477984 & 184 \\
& Math75 & 8 & 1 & 26.6 & 4.8 & 31 & 54885 & 239 \\
& Math82 & 1 & 1 & 179.7 & 39.7 & 21 & 643677 & 103 \\
& Math94 & 14 & 1 & 20.5 & 0.5 & 1 & 171 & 334 \\
& Math02 & 23 & 1 & 95 & 0.8 & 1 & 12459 & 323 \\
& Math05 & 13 & 4 & 24.6 & 5.9 & 63 & 1743 & 418 \\
& Math11 & 1 & 1 & 34 & 7.8 & 2 & 53576 & 28 \\
& Time19 & 5 & 3 & 54.1 & 27.8 & 18 & 6546 & 345 \\ \cline{2-9}
& \textbf{Median} & 2 & 1 & 28.85 & 7.05 & 19.5 & 24089.5 & 130 \\
& \textbf{Average} & 4.9 & 1.4 & 40.8 & 11 & 27.6 & 191582.2 & 166.3 \\
& \textbf{StdDev} & 6.4 & 0.9 & 43.3 & 13.2 & 35.2 & 387814.5 & 131.7 \\ \hline \hline

\multirow{16}{*}{No}
& Chart03 & 0 & -1 & 43.1 & 0 & 0 & 0 & 142 \\
& Chart09 & 3 & -1 & 16.2 & 0 & 0 & 0 & 77 \\
& Lang21 & 2 & -1 & 10.6 & 0 & 0 & 0 & 159 \\
& Lang24 & 4 & -1 & 9.7 & 0 & 0 & 0 & 116 \\
& Lang61 & 0 & -1 & 30.9 & 0 & 0 & 0 & 316 \\
& Math41 & 8 & -1 & 7.7 & 0 & 0 & 0 & 124 \\
& Math58 & 0 & -1 & 13.4 & 0 & 0 & 0 & 32 \\
& Math96 & 8 & -1 & 24 & 0 & 0 & 0 & 357 \\
& Math104 & 0 & -1 & 2.6 & 0 & 0 & 0 & 0 \\
& Math105 & 10 & -1 & 4.1 & 0 & 0 & 0 & 34 \\
& Math27 & 0 & -1 & 30.8 & 0 & 0 & 0 & 473 \\
& Math30 & 49 & -1 & 54.3 & 0 & 0 & 0 & 411 \\
& Math32 & 2 & -1 & 18.1 & 0 & 0 & 0 & 104 \\
& Math34 & 0 & -1 & 5.3 & 0 & 0 & 0 & 0 \\
& Time04 & 49 & -1 & 78.1 & 0 & 0 & 0 & 316 \\
& Time16 & 1 & -1 & 21.1 & 0 & 0 & 0 & 57 \\ \hline

\end{tabular}}
\label{table2}
\end{table}

The results for QuixBugs are shown in Table~\ref{table1}.
From left to right, the first column shows whether the error was correctly repaired or not. 
The second column shows the error names along with statistics for errors correctly repaired. 
The \textit{\#Results} column shows the number of repairs returned by the back end.  
The \textit{Correct Rank} column gives the rank of the correct repair if 
it was found, and -1 if it was not.
The \textit{Total Time} column gives the total time in seconds for all repairs to be found.  
The \textit{Fix Time} column shows the time in seconds for the correct repair to be reported and viewed.
The \textit{Fix Count} and \textit{SEEDE Count} columns show the number of repairs validated and total number of
instructions SEEDE executed  before processing this correct repair, measures of the amount of work
done before finding the correct repair. Since ROSE discards repairs after an instruction count 
is exceeded to provide faster results, these 
indicate what might happen if ROSE stopped validating repairs earlier.
The \textit{\#Checked} column gives the number of repairs on which validation was run.

These results demonstrate that ROSE is practical and useful for simple programs.
ROSE successfully found repairs for 17 of the 40 QuixBugs problems with a median total time of $\bettersim$5 seconds.
Moreover, the maximum time to produce a successful repair (i.e., the Fix Time) is under 4 seconds.
Of the repairs generated, 14 were the top ranked repair, two of them were ranked second, and one ranked fourth.
Previous studies with QuixBugs indicated 
that different techniques repair 1-6 errors with a 5-minute timeout~\cite{ye2021comprehensive} or 9-11 errors 
with a median time of 14 to 76 seconds~\cite{asad2020impact}.  Thus ROSE was able to repair significantly
more errors in significantly less time.  

The results for Defects4J are shown in Table~\ref{table2}.  ROSE was able to 
find the correct repair in 16 of the 32 errors.  For 12 of these, the correct repair was top ranked.  
Two other correct repairs were second ranked, and then one correct repair was third ranked and one was fourth. 
The median time used by ROSE to find all repairs for a
given error was $\bettersim$29 seconds; the maximum time to find the correct repair 
was $\bettersim$40 seconds and the median time for this was $\bettersim$7 seconds. 
Compared to 22 APR techniques whose repairs were reported for the 32 errors, we found that 
ROSE repaired more errors than 20 techniques including ACS~\cite{xiong2017precise},
LSRepair~\cite{liu2018lsrepair},
CapGen~\cite{wen2018context}, SimFix~\cite{jiang2018shaping}, 
sharpFix~\cite{xin2019better}, SequenceR~\cite{chen2019sequencer}, and 
TBar~\cite{liu2019tbar}, which, according to their published results, correctly 
repaired 4, 6, 9, 7, 11, 9, and 14 errors respectively. The only two techniques 
outperforming ROSE are Hercules~\cite{saha2019harnessing} and CURE~\cite{jiang2021cure}, which 
repaired 18 and 17 errors respectively.  
Among the 22 APR techniques, 12 reported the time 
(median or average) used to produce a repair.
LSRepair is the most efficient with its median repair time being 36 seconds. 
For all other techniques, the repair time is over 4 minutes.
These results demonstrate that ROSE is practical even for relatively complex systems
assuming the developer is currently debugging and the repair is relatively simple. 

\subsection{Discussion}

ROSE worked quite well and fast for the simple errors of QuixBugs.  It did 
significantly better than previous techniques in finding correct repairs.  
This was due to several factors, including ROSE’s ability to handle infinite 
loops, using a combination of repair techniques, 
suggesters that are designed to handle the common programming problems 
that arise in student programs, and the fact that ROSE was provided with more 
input than just whether a test succeeded or failed.  ROSE did not handle 
cases where the errors required multiple changes and missed most of the repairs that 
required inserting new code rather than repairing an existing line.  This was expected 
as the typical code that needs to be added is very specific to the 
program and is unlikely to be found by a simple repair tool.  This
is also in line with the goals of ROSE which were to fix simple errors, not to handle complex cases.

ROSE did reasonably well on the Defects4J errors it was tested on, but not as well 
if one considers the whole Defects4J corpus with several hundred errors.
ROSE is designed to handle relatively simple 
errors (hence the Overt in its acronym). Most of the Defects4J errors require 
sophisticated repairs that involve understanding the semantics of the code, often multiple
insertions and deletions.  Moreover, in 
several of the tests used in the corpus, for example those in \textit{Closure}, the 
failing assertion is far from the error line, violating ROSE's assumptions 
that the developer has done some preliminary work to narrow down the problem.  
Three of the reported tests 
failed because they expected the code to throw a particular 
exception.  While this occurs in testing complex libraries and occurs in the Defects4J corpus, which is 
largely based on such libraries, it is not
particularly common in general systems.  The current version of 
ROSE does not understand this as a problem, and 
ROSE is unable to find solutions in these cases.  

ROSE’s time in repairing the 
Defects4J errors is not as good but is still much better than existing systems
and would still be practical in an interactive development environment. 
The repair time can be dominated by the length of the execution from the starting point, 
the number of potential error lines, or the number of repairs that need to be validated, 
and differed with different problems.

From these experiments, one can see the trade-off ROSE is trying to achieve between the time
taken to produce results and the number and accuracy of the results reported.  To be useful in
an interactive setting, ROSE needs to provide the developer with results reasonably quickly.  This
has to at least be faster than the developer thinks they would take to identify and fix the problem.
If ROSE stopped reporting results sooner, some of the repairs found would not have been reported. 
Similarly, if ROSE would spend more time finding results, it would be able to find additional correct
repairs.  Our initial experiments showed at least one additional correct repair would be returned for 
both the QuixBugs and Defects4J errors that were considered if ROSE had more time.
The results given show that overfitting is not a major problem so far as most of the correct repairs are ROSE's
highest priority.   
ROSE assumes the developer will choose the correct repair, 
so having it in the top-5 listed is probably sufficient. 
This is similar to Quick Fix where the desired repair is sometimes not the first one listed.

\subsection{Threats to Validity}

While these experiments show that ROSE has promise, they are not definitive.  
First, ROSE makes different assumptions than the APR systems we use for comparison.
ROSE uses 
information beyond a test suite.  To the extent that we provided such information, 
for example
setting a break point in an infinite loop or identifying a particular variable value 
as wrong, the results obtained by ROSE are not directly comparable.  
However, disregarding this information and using ROSE only on the 
failing test assertion would not be a typical use of ROSE or a 
typical interactive situation.  

Second, current implementation of ROSE is only a prototype.  We are still working 
on adding new suggesters, fixing priority assignments, improving fault localization, 
and tweaking the validation algorithms.  New suggesters might find additional 
correct repairs but could cause the overall time to find a repair to increase; 
such suggesters might also find overfitted repairs more frequently than existing methods.  
Moreover, the current implementation has some limitations such as the inability to handle
detailed Java reflection.

We are planning on continuing to develop ROSE to the point where it can become an
everyday part of an IDE.  At that point we plan to conduct user studies and a long-term
in situ study of its use.



\section{Conclusion}
\label{sec:conclusion}

ROSE is a new tool designed for automatically repairing semantic errors that arise during 
debugging within an IDE.
ROSE does not assume an extensive test suite but depends on a quick error description and the 
current execution environment for repair. It uses a flow-analysis based technique
for fault localization, a combination of three types of approaches to make repairs, 
and a technique based on execution comparison for repair validation.
Our evaluation demonstrated that ROSE was useful for repairing problems in simple programs,
and also fast and capable of handling complex systems.  It also shows that an extensive
test suite is not required to do program repair.
A video showing ROSE in action can be seen at \url{https://youtu.be/GqyTPUsqs2o}.

ROSE is an available open-source project.  The experimental data including a 
code snapshot are available from \url{https://sites.google.com/view/roserepair}.  
ROSE currently 
runs with the Eclipse environment and is implemented as a separate process so it 
could be ported to other IDEs.  The tools ROSE uses, FAIT and SEEDE, are also open source and available.
ROSE's user interface 
was developed for and is distributed with the open source Code Bubbles IDE.


\clearpage

\bibliographystyle{plain}
\bibliography{paper}

\end{document}